\documentclass[a4paper,11pt]{article}

\usepackage[utf8]{inputenc}   
\usepackage{amsmath}          
\usepackage{amssymb}          
\usepackage{graphicx}         
\usepackage{color}
\usepackage{hyperref}         
\usepackage{geometry}         
\usepackage{natbib}           
\usepackage{authblk}          
\usepackage{appendix}
\newcommand{\xmm}{\textit{XMM-Newton }}
\title{{\it XMM-Newton} high-resolution spectroscopy of EXO 0748--676 after its re-emergence from a long quiescence}
\author[1]{Sayantan Bhattacharya\thanks{
sayantan34@gmail.com}}
\author[1]{Sudip Bhattacharyya\thanks{
sudip@tifr.res.in}}
\author[1]{Gargi Shaw\thanks{
gargishaw@gmail.com}}
\affil[1]{Department of Astronomy and Astrophysics, Tata Institute of Fundamental Research, 1 Homi Bhabha Road, Colaba, Mumbai 400005, India}
\date{November 14, 2024}

\begin{document}

\maketitle

\begin{abstract}
EXO 0748--676 is a well-studied, high-inclination, dipping, and eclipsing neutron star low-mass X-ray binary that has recently emerged from 16 years of quiescence into a new outburst. We present results from a 55.5 ks of \textit{XMM-Newton} observation, focusing on high-resolution spectroscopy with the same instrument (Reflection Grating Spectrometer) that produced significant insights during the previous outburst.
The \textit{XMM-Newton} European Photon Imaging Camera light curve reveals a type I X-ray burst which leads to a corresponding optical burst by three seconds. 
To understand the effects of the burst on the ionization structure, the data are divided into burstless, pre-, and post-burst spectra, with additional analysis for dip and non-dip phases. The primary spectral feature in all phases is a broad O VII recombination line, accompanied by velocity-broadened O VIII, N VII, and Ne IX lines. Notably, the Ne IX line shows different ionization states for pre-burst (11.65 \AA{}) and post-burst (13.56 \AA{}) phases, while the dips also substantially affect spectral lines.
The current outburst mirrors many traits from the earlier one, such as a similar spectral state, plasma components with similar ionization structure, and spectral features from the same elements, implying a stable long-term accretion behavior across outbursts. 
\end{abstract}

\twocolumn
\section{Introduction} \label{sec:intro}
EXO 0748--676 is a neutron star (NS) low-mass X-ray binary (LMXB) system, that has been studied extensively since its discovery in 1985 \citep{parmar1986discovery}. 
In such a system, an NS accretes matter from a low-mass companion star.
The appearance of thermonuclear (type I) X-ray bursts confirmed the compact object type as a neutron star \citep{gottwald1987properties}. 
EXO 0748--676 is a prolific burster 
with the recurrence time of $\sim 3$ hr to 10--12 minutes, and bursts can occur as singlets, doublets, and triplets \citep{boirin2007discovery}. 
The bursts typically exhibit a linear rise followed by an exponential decay profile, with the duration of $\sim 15-177$ s and the characteristic decay time in the range of $\sim 14-57$ s \citep{boirin2007discovery}. 
Two burst oscillation features \citep[e.g., ][]{bhattacharyya2021nuclear} at 45Hz \citep{villarreal2004discovery} and 552 Hz \cite{galloway2010discovery} were also observed.
The source has been associated with an optical counterpart UY Vol \citep[binary orbital period $P_{\rm orb} \approx 3.82$ hr; ][]{wade1985optical}. 
X-ray eclipses and dips are observed in light curves \citep{homan2003xmm},
suggesting a high inclination \citep[75$^\circ$--82$^\circ$; ][]{parmar1986discovery} of the system. The eclipses are of $\sim8.3$ min with a typical ingress/egress time of 8--10 s. 
The intensity dips are the result of absorption by photoionized clouds 
situated above the accretion disk. 
During the intensity dips the X-ray spectrum hardens, as is expected from enhanced photoelectric absorption \citep{parmar1986discovery}. 
Another spectral characteristic of EXO 0748--676 is a soft excess in the 8--30{\AA} band, observed with {\it ASCA} and {\it ROSAT} \citep{thomas1997discovery,schulz1999rosat}. \citet{homan2003xmm} reported energy-dependent variations in the light curve of the system. 
The eclipses are more prominent in the hard band (2--10 keV) and sometimes not at all observed in the soft band \citep{homan2003xmm}. 
These authors also showed that dipping is more evident in the soft band (0.3--2 keV). 
They explained the spectral state evolution of this system as a result of variable absorption covering two emission components, where the extended component is softer than the central source.

The high inclination of EXO 0748--676 implies that the observer's line of sight is close
to the accretion disk, and hence, the source should be ideal for probing the structures above the disk using observed spectral lines.
High-resolution spectra from both \textit{XMM-Newton} Reflection Grating Spectrometer (RGS) and \textit{Chandra} High Energy Transmission Grating (HETG) were used earlier to 
probe the spectral lines and other properties of this system. 
\citet{cottam2001high} reported the RGS spectral analysis to compare the spectra from low emission period, rapid variation period, and bursts. 
They found broadened recombination lines from N VII, O VII, O VIII, Ne IX, and Ne X (velocity $\sim 10^3$ km s$^{-1}$), and the broadening varied with the ionization parameter. 
The K edges of the O VII and O VIII were also detected during the rapid variation phases. 
\citet{cottam2002gravitationally} also reported gravitationally redshifted ($z=0.35$) absorption features (Fe XXV n=2-3, Fe XXVI n=2-3, O VIII n=1-2). 
These features are crucial in determining NS parameters like radius-to-mass ratio \citep{bhattacharyya2006shapes} and probing the superdense matter of the NS core \citep{bhattacharyya2010measurement}.
However, while these features were detected in the 2000 observations of the bursts, they were not found in the later 2003 observations \citep{cottam2008burst}.

\citet{jimenez2003discrete} used 
\textit{Chandra}/HETG spectroscopy to reveal the nature of photoionized plasma orbiting the neutron star. They detected the lines reported by \citet{cottam2001high} along with Mg XII Ly$\alpha$ and Mg XI lines in the summed HEG and MEG spectrum. Their findings indicate the existence of a photoionized plasma located in the outer radius of the accretion disk and the plasma is located at a height of $\sim0.2$r above and below the disk midplane.

In a later work using archival {\it XMM-Newton}/ RGS data, the persistent and dip spectra of this system were studied in detail \citep{van2009properties}. 
They inferred that the continuum was absorbed by both collisionally ionized and photoionized plasma. The photoionized absorber shows significant enhancement in column densities during the dips. They mainly detected the O VII triplet and the O VIII Ly$\alpha$ lines. The FWHM for O VII varies from 0.25{\AA} to 0.5{\AA}, and the same for the  O VIII Ly$\alpha$ line varies between 0.05{\AA} to 0.35{\AA}, which is similar to \citet{cottam2001high} observations. Notably, \citet{jimenez2003discrete} determined smaller FWHM values using \textit{Chandra} than those by both \citet{cottam2001high,van2009properties}. 

\citet{psaradaki2018modelling} used 11 archival \xmm observations to study the eclipse spectrum. They detected the same O VII, O VIII, Ne IX, and N VII lines. Their results of density diagnostics using the O VII triplet lines suggest the existence of two photoionized plasma components with different ionization parameters ($\log \xi \sim 2.3$ and $1.3$). The second component covers a smaller fraction of the source but exhibits an inflow velocity.

After $\approx 23$ years of its first detection, EXO 0748--676 faded into quiescence in 2008 \citep{hynes2009quiescent}.
However, it has recently returned to an active accretion phase after $\approx 16$ years \citep[e.g., see ][]{baglio2024lco,bhattacharya2024xmm,buisson2024updated}. 
Given the utility of spectral lines from EXO 0748--676 for probing the accretion structure and other properties of this source, and generally of NS LMXBs, it is immensely interesting to check if similar lines appear during the new outburst, and if yes, to compare them with the lines of the previous outburst. 
Such a study and comparison is best done using the new {\it XMM-Newton}/RGS data because a huge amount of the same instrument's data from the previous outburst of the source exists.
With a preliminary analysis of the current {\it XMM-Newton}/RGS data, we reported an oxygen emission feature \citep{bhattacharya2024xmm}.
Motivated by this finding, here we report the analysis of the entire available {\it XMM-Newton} data during the current outburst of EXO 0748--676.

This work is structured as follows. In section \ref{methods}, observation and data analysis methods are explained. 
Section \ref{epicburst} presents the \textit{XMM-Newton} European Photon Imaging Camera (EPIC) light curve and a burst, and section \ref{rgs} presents the results of high-resolution spectral analysis. 
In section \ref{disc}, we discuss the results, and in section \ref{conc}, we summarize this work and motivate further research.

\section{Observation and data analysis}\label{methods}
The observation of EXO 0748--676 with \xmm was performed on June 30, 2024, for an exposure of 55.5 ks. 
This work mainly focuses on the high-resolution spectrum obtained using the RGS instrument. In the present work, the EPIC data are used to create Good Time Intervals (GTIs) and study the type I burst and continuum spectrum.

The RGS spectrum covers a wavelength range of 5--35 {\AA} with a resolution of 0.05 {\AA} (mostly constant over the whole wavelength range). 
The raw \textit{XMM-Newton} data are reduced using SAS 21.0, and specifically, the RGS 1 and 2 event files, spectra (both 1st and 2nd order), and light curves are generated using \texttt{rgsproc}. 
Then the RGS 1 and RGS 2 spectra of the same order are combined using \texttt{rgscombine}. This combined spectrum is finally analyzed using \texttt{XSPEC 12.14}\footnote[1]{\texttt{XSPEC 12.14} \citep{1996ASPC..101...17A} is part of the \texttt{HEASoft} software package, available at \url{https://heasarc.gsfc.nasa.gov/xanadu/xspec/}.}. For line detection, first, we identify prominent lines and fit them with Gaussian (\texttt{gauss}) emission function in \texttt{XSPEC}.
For other less significant features, we look for lines corresponding to elements detected in earlier studies \citep{cottam2001high,van2009properties,psaradaki2018modelling}. This is also confirmed using plasma modeling with CLOUDY \citep{2023RMxAA..59..327C}
discussed in more detail in section~\ref{Sec:CLOUDY}. 
The significance of the lines is also tested by scanning the entire spectra and fitting them with Gaussian emission and using the F-test with two models, just fitting a continuum and another adding a Gaussian line to the continuum model \citep{pinto2016resolved}. The detailed description and figure are in Appendix~\ref{gauss_scan}. 

The EPIC-PN and MOS event files are generated using \texttt{epproc} and \texttt{emproc}. Then the \texttt{evselect} routine is used to create PN, MOS1, and MOS2 source and background light curves and spectra. All the light curves are background subtracted and combined using \texttt{lcmath} to obtain the net source light curve.

The EPIC light curve is used to identify the burst duration and midpoint. Then \texttt{tabgtigen} is used to create GTIs for pre-burst (836169426 s to 836205700 s Mission Elapsed Time (MET)), post-burst (836206600 s to 836224960 s MET), and burstless duration (54.6 ks, combination of the pre- and post-burst duration). These GTI files are again used with \texttt{rgsproc} to create pre-burst, post-burst, and burstless total spectral data products. These first-order spectra are then grouped to contain at least 5 counts in a bin using \texttt{grppha}.

\section{EPIC light curve, type I X-ray burst, and hardness}\label{epicburst}  

\begin{figure*}[h]
    \centering
    \includegraphics[width=0.95\textwidth]{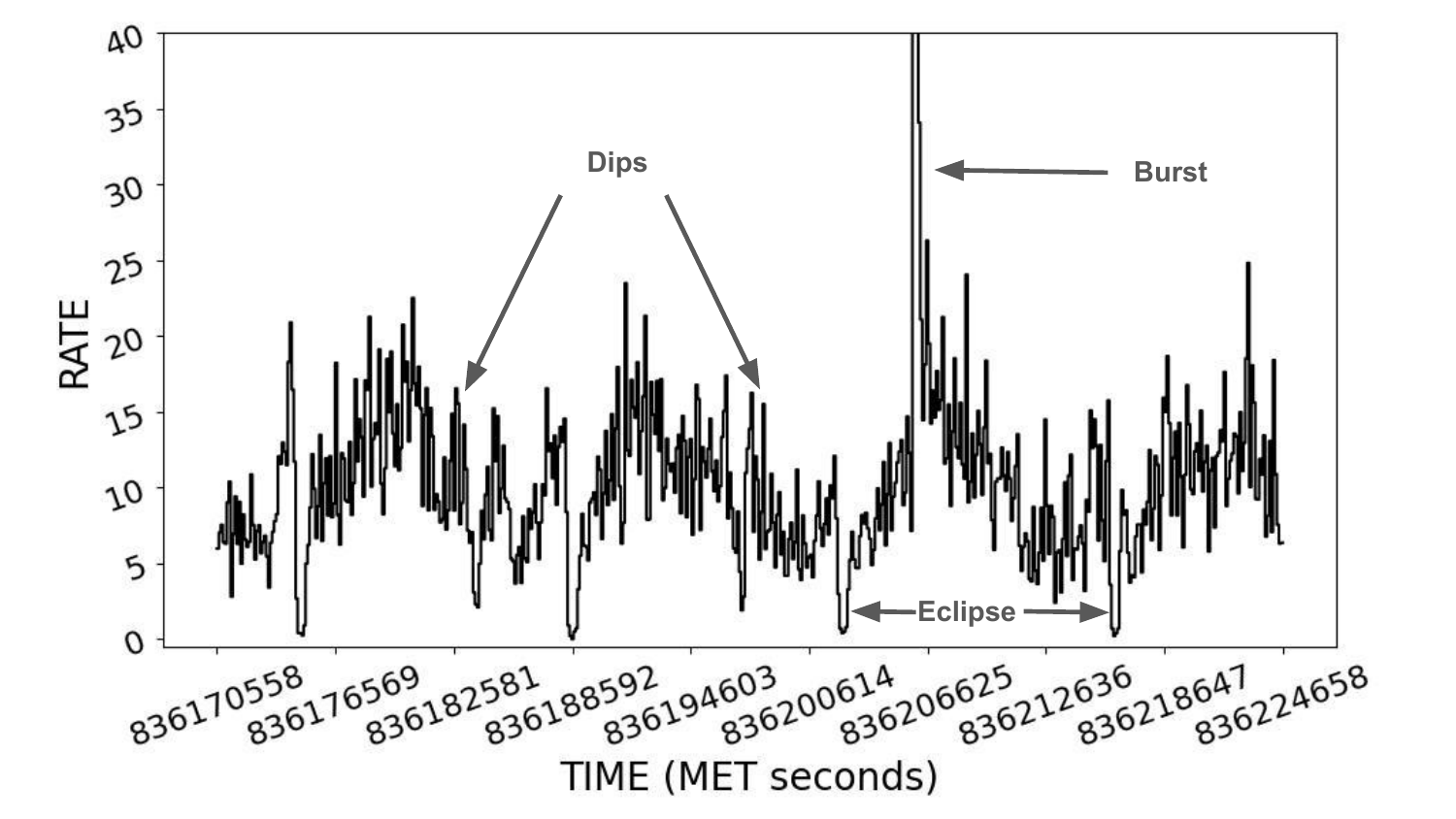}
    \caption{\xmm MOS light curve of EXO 0748--676 showing dips, eclipses, and a type I X-ray burst at MJD 60492.3097 (MET 836205954 s) (see section~\ref{epicburst}).}
    \label{fig:epiclc}
\end{figure*}

The PN (timing mode), MOS1, and MOS2 (imaging mode) net light curves of this observation are added together to create a broadband (0.2--10 keV) EPIC light curve. A type I X-ray burst is observed in the combined as well as individual time series (Figure~\ref{fig:epiclc}). 
The peak time for the burst is MJD 60492.3097. 

The spectral state of the source is necessary to understand and compare spectral features with past/future observations (detailed in Appendix \ref{hardness}). The eclipses in the broadband (0.3--10 keV) light curve are not as prominent as those observed in the hard band (2--10 keV) light curve. This is also seen in previous studies where the eclipses were absent in the soft but present in the hard band light curves \citep[see][for a detailed discussion]{homan2003xmm}. 

\begin{figure*}[h]
    \centering
    \includegraphics[width=\textwidth, height=1.2\textwidth]{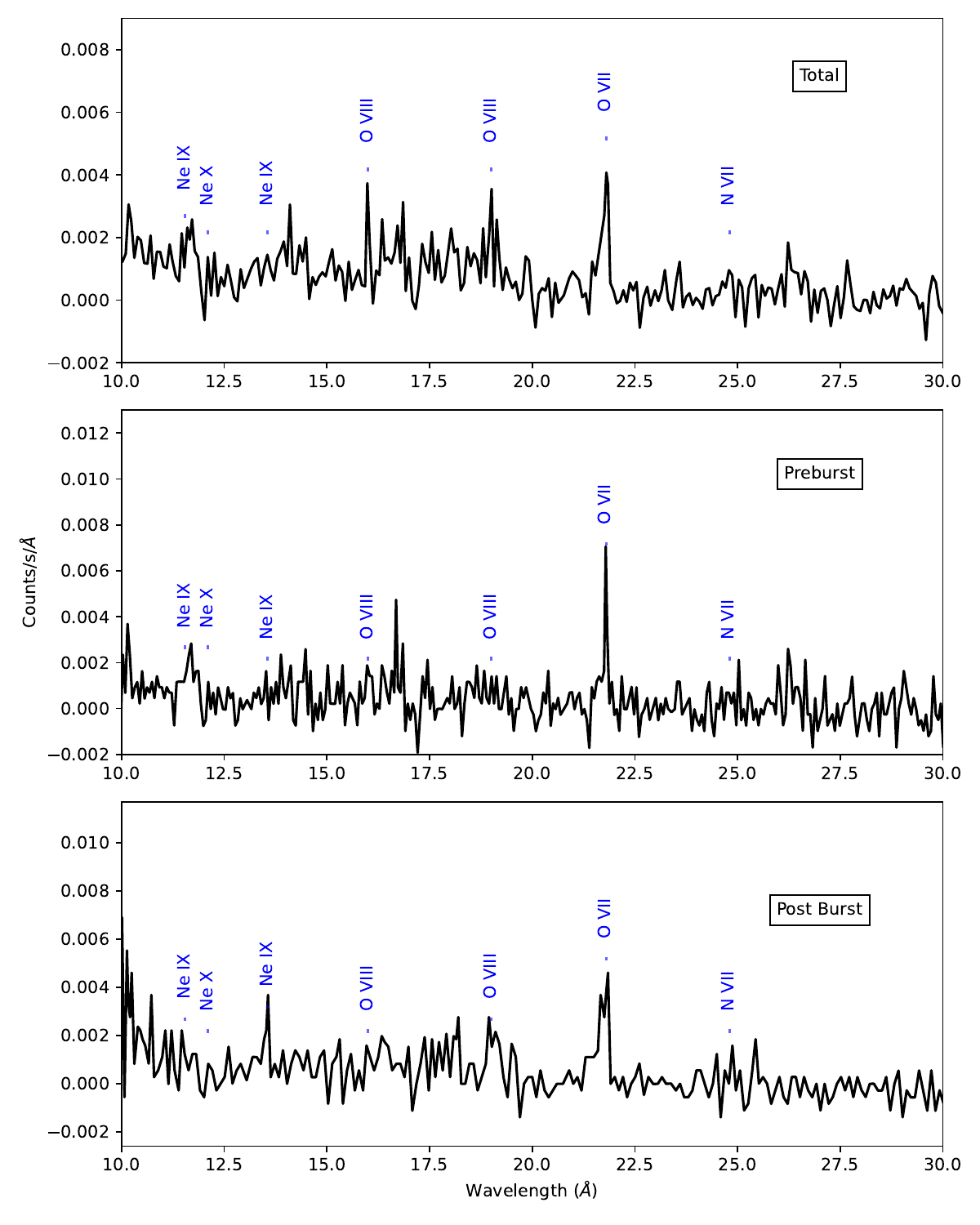}
    \caption{\textit{XMM-Newton}/RGS spectra for total observation, pre-burst, and post-burst phases of EXO 0748--676. The O VII intercombination line is evident in all three phases. The broad Ne IX feature changes its ionization state from pre-burst spectrum to post-burst spectrum (see section~\ref{rgs}).}
    \label{fig:rgs_plot}
\end{figure*}

\begin{table*}[tbh]
    \centering
    \caption{Spectral line (Gaussian) best-fit parameters from total (excluding burst), pre-burst, post-burst, dip, and non-dip \xmm RGS spectra of EXO 0748--676 (see section~\ref{rgs_prepost}). 
    }\vspace{0.3 cm}
    \begin{tabular}{c c c c c c}
         Line ID & Line Center & FWHM & FWHM & \texttt{gauss} norm & Detection  \\ 
                 & (\AA) & (\AA) & ($\times 10^3$~km s$^{-1}$)& ($\times10^{-4}$~photon/cm$^2$/s) & \% Probability\\
         \hline \hline
          & &Total & \\ \hline
         Ne IX (1-3)  &  11.65$\pm$0.03 & 0.28$\pm$0.06& 7$\pm$1 & {0.4$\pm$0.1} & {90}\\
         O VIII & 18.99$\pm$0.03  & 0.21$\pm$0.08 & 3$\pm$1 & {0.8$\pm$0.2} & {90}\\
         O VII & 21.79$\pm$0.02 & 0.24$\pm$0.03 & 3.3$\pm$0.4 & {1.7$\pm$0.2} & {$>$99}\\
         N VII & 24.81$\pm$0.04 & 0.17$\pm$0.05 & 2.0$\pm$0.6 & {0.22$\pm$0.09} & {88}\\ \hline
         & &Pre-Burst & \\ \hline
         Ne IX (1-3)  & 11.67$\pm$0.06 & 0.30$\pm$0.09& 7$\pm$2 & {0.4$\pm$0.1} & {91}\\
         O VIII & 18.95$\pm$0.09 & 0.2$\pm$0.1 & 3$\pm$1 & {0.2$\pm$0.1} & {83}\\
         O VII & 21.80$\pm$0.02 &0.15$\pm$0.03 & 2.1$\pm$0.4 & {1.8$\pm$0.4} & {95}\\
         N VII & - & - & - & - & - \\ \hline
         & &Post-Burst & \\ \hline
         Ne IX (1-2)  & 13.56$\pm$0.07 & 0.3$\pm$0.1 & 7$\pm$2 & {0.51$\pm$0.06} & {$>$99}\\
         O VIII & 19.1$\pm$0.1 & 0.3$\pm$0.1 & 5$\pm$1 & {0.6$\pm$0.1} & {95}\\
         O VII &  21.78$\pm$0.04 & 0.20$\pm$0.09 & 2.7$\pm$0.8 & {1.4$\pm$0.5} & {$>99$}\\
         N VII & - & - & - & - & - \\ \hline
         & &Dip & \\ \hline
         Ne IX (1-3)  & 11.65$\pm$0.05 & 0.3$\pm$0.1 & 7$\pm$2 & {0.5$\pm$0.1} & {95}\\
         Ne IX (1-2) & 13.53$\pm$0.03 & 0.11$\pm$0.05 & 2.3$\pm$0.8 & {0.27$\pm$0.06} & {88}\\
         O VIII & 19.01$\pm$0.04 & 0.3$\pm$0.1 & 5$\pm$1 & {0.4$\pm$0.1} & {95} \\
         O VII  &  21.79$\pm$0.02 & 0.14$\pm$0.07 & 1.9$\pm$0.9& {1.26$\pm$0.08} & {$>$99}\\
         N VII  & - & - & - & - & -\\ \hline
         & &Non-Dip & \\ \hline
         Ne IX (1-3) & 11.66$\pm$0.04  & 0.26$\pm$0.09 & 6$\pm$1 & {0.38$\pm$0.04} & {$>99$} \\
         O VIII & 19.01$\pm$0.01 & 0.08$\pm$0.03 & 1.3$\pm$0.4 & {0.8$\pm$0.3} & {98}  \\
         O VII & 21.80$\pm$0.03 & 0.29$\pm$0.06 & 3.9$\pm$0.8 & {1.3$\pm$0.4} & {$>$99} \\
         N VII & - & - & - & - & - \\
    \end{tabular}
    
    \label{tab:line_table}
\end{table*}

\section{RGS high resolution \\ spectrsocopy}\label{rgs}
The high-resolution spectrum of the entire observation reveals various spectral features related to different physical properties of the system. The RGS data of EXO 0748--676 were utilized in multiple previous studies for this purpose as summarized in section~\ref{sec:intro}.

\subsection{Pre-, post- and burstless total spectra}\label{rgs_prepost}
To better understand the evolution of the spectral characteristics, the entire data are divided into three parts: pre-burst, post-burst, and the entire observation excluding the burst (see section~\ref{methods} for GTI durations). The spectrum for each phase is then fitted with the \texttt{pcfabs*cutoffpl} \texttt{XSPEC} model, with the initial values from fitting the same model to the MOS CCD spectrum (see Appendix~\ref{mos_spec}). Then the data is fitted with Gaussians (\texttt{gauss} in \texttt{XSPEC}) for spectral lines. In order to confirm the line detections, the entire spectrum is scanned and fitted using Gaussian emission and using F-test with two models, just fitting a continuum and another adding a Gaussian line to the continuum model. This method is also followed by \citet{psaradaki2018modelling} and \citet{pinto2016resolved}. This method is described in Appendix~\ref{gauss_scan} with the corresponding figure (Fig.~\ref{gaus_scan}). The line properties and their detection probability are listed in Table~\ref{tab:line_table}.  

The broad O VII emission feature (21.8 {\AA}) is evident in all phases of the spectra. This feature is dominated by the O VII, n=1-2 intercombination line (x, y). This is also accompanied by the O VII resonance absorption line (w) at 21.6 {\AA}. Other notable features include Ne IX, O VIII Ly$\alpha$, and N VII emission, as well as the O VII absorption edge from circumstellar contributions (see Figure~\ref{fig:rgs_plot}).

In addition to these spectral features due to circumstellar ionized material, \citet{cottam2002gravitationally} also detected absorption from the neutron star photosphere at 13.8 {\AA}, 25.2 {\AA}, and 26.4 {\AA}. In this case, the spectra are noise-dominated after 23 {\AA}, making it hard to detect the 25.2 or 26.4 {\AA} feature. We do not detect an absorption feature near 13.8 {\AA}. However, a broad Ne IX emission line is detected at 13.56 {\AA} with the possibility of line blending, as Ne IX triplet is often found to be blended with weaker Fe lines \citep{pinto2013interstellar}.

The three different segments are compared to each other, and most of the spectral features appear in all of them, with a few exceptions.
The O VII intercombination line is detected in all three of these, as are the O VIII Ly$\alpha$ line and the O VII absorption edge. 
However, in the pre-burst spectrum, an emission feature is observed at 11.67 {\AA}, which is not detected in the post-burst phase. But, another emission line at 13.56 {\AA} appears in the post-burst phase, which is not detected in the pre-burst phase. 
Using AtomDB \citep{smith2014atomdb} to look for lines in these two periods, it is found that the Ne IX 11.55 {\AA}, n=1-3 and the Ne IX 13.55 {\AA}, n=1-2 lines have the highest emissivities. 
Hence, it can be inferred that the Ne IX ion is detected for these two different transitions during the pre-and post-burst phases. 
Although these lines show large FWHM values, similar broad lines were also reported in \citet{van2009properties}.
The N VII 24.8 {\AA} emission line is only detected in the total spectrum and not in the pre-burst spectrum. The post-burst spectrum is noise-dominated in this wavelength range.

Finally, comparing with previous high-resolution spectral studies, we find that this observation has more similarity with the  \citet{cottam2001high} previous outburst results because the burst is excluded from the total spectrum in the present work. 
Later, \citet{cottam2002gravitationally} and \citet{cottam2008burst} used only the stacked burst spectra, a similar study cannot be carried out with this observation exhibiting a single burst. 
Hence, to better probe the current outburst and separately analyze the persistent and burst (early and late phases) emissions, more high-resolution observations are required.   

\subsection{Dip and non-dip spectra}\label{dip_non}
\begin{figure*}[h]
    \centering
    \includegraphics[width=0.96\textwidth]{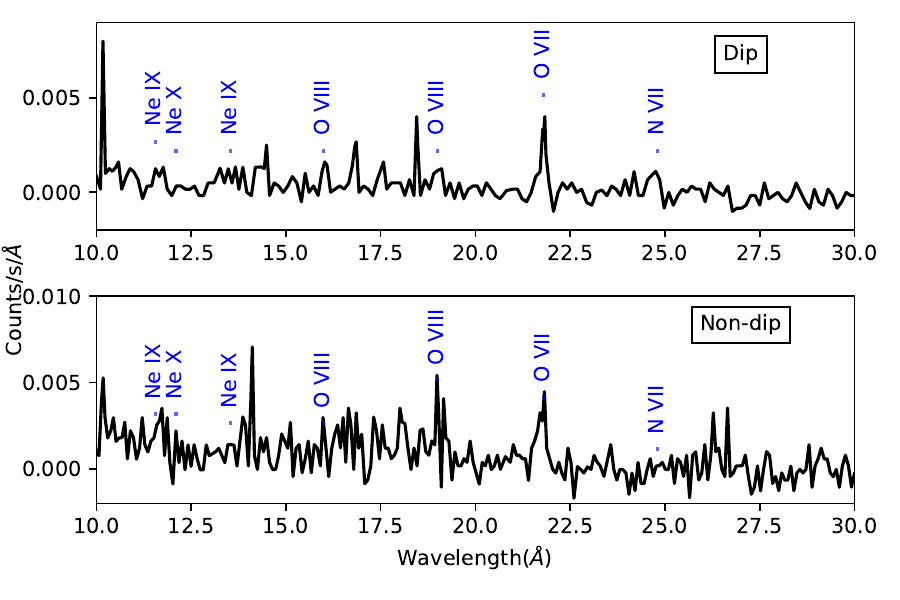}
    \caption{XMM-Newton/RGS spectra of dip (top) and non-dip (bottom) phases of EXO 0748-676.
    This figure shows the effects of intensity dips on spectral lines (see section~\ref{dip_non}).}
    \label{fig:dip-nodip}
\end{figure*}

High-inclination LMXB systems, such as EXO 0748--676, show intensity dips \citep{homan2003xmm}. 
We find such dips from the current \xmm observation, and the RGS spectra created for dip and non-dip periods for this source are compared (Figure~\ref{fig:dip-nodip}). The line profile parameters and detection probability are obtained using the same method as mentioned in section~\ref{rgs_prepost}, and given in Table~\ref{tab:line_table}. 

Both the dip and non-dip spectra exhibit the strong O VII intercombination line (21.8\AA) and the O VIII Ly $\alpha$ emission line (19\AA). The Ne IX lines are also detected, the dip spectrum shows both the Ne IX lines at 11.65 and 13.5{\AA}, whereas the non-dip spectrum doesn't have the 13.5{\AA} line. The N VII (24.8\AA) line is not detected in the non-dip phase and the wavelength region of N VII (24.8\AA) emission in the dip phase is noise-dominated.

The line characteristics show some difference between the dip and non-dip periods. 
While the O VII line is detected in both phases, the line intensity of the O VIII Ly $\alpha$ emission line is higher in the non-dip phase.  

\subsection{CLOUDY photoionization modeling}\label{Sec:CLOUDY}
As mentioned above, multiple emission lines are observed in the high-resolution spectra of EXO 0748--676. 
In order to understand the origin of these lines, we use the spectral synthesis code CLOUDY 23.01 \citep{2023RMxAA..59..327C}. CLOUDY is based on the microphysics of the ionizing environment; it calculates the ionization, excitation, energy balance, chemistry, and radiation transport simultaneously and self-consistently \citep{2013RMxAA..49..137F, 2022ApJ...934...53S, 2023RNAAS...7..153S}. As a result, it predicts the underlying physical conditions and the emitted/absorbed spectra over the entire electromagnetic range. Earlier, \citet{2019MNRAS.486..195S} had used CLOUDY to model four NS LMXB spectra. In the present paper, we have used CLOUDY to model EXO 0748--676's spectrum for the first time. We consider two different incident radiations irradiating the hot gas cloud, i.e., the hot plasma. One is a black body radiation with temperature 10$^7$ K and luminosity 10$^{36.5}$ erg s$^{-1}$. The other one is a power law radiation ($\nu ^{\alpha}$) with $\alpha$ = -1.44, and luminosity 10$^{35}$ erg s$^{-1}$. The continuum model values are chosen by fitting the observed EPIC spectrum by a blackbody and power law model. The luminosity, hydrogen number density $n_{\rm H}(r_{0})$(cm$^{-3}$), inner radius r$_0$ (cm), and thickness (cm) are free parameters. The density profile is the same as \citet{2019MNRAS.486..195S}, i.e n(r) = n$_0$(r$_0$)$\times$(r/r$_0$)$^{-2}$, facing from the ionizing source. 

Our best model parameters and line ratio predictions are given in Appendix~\ref{cloudyfigtab}, Table \ref{tab:table1}, and Table \ref{tab:table2}. The existence of a high-density plasma can be observed, a similar high density was estimated by \citet{psaradaki2018modelling}. Initially, we considered solar abundances \citep{2010Ap&SS.328..179G} for all the elements. The predicted column densities of O VII, O VIII, N VII, and Ne IX are $4.82 \times 10^{19}$, $4.02 \times 10^{19}$, $6.95 \times 10^{18}$, and $9.98 \times 10^{18} \ \mathrm{cm}^{-2}$, respectively. The predicted electron temperature averaged over radius is 2.13$\times$10$^5$ K. 
This predicted electron temperature is derived from the heating and cooling balance consisting of various physical processes \citep{2013RMxAA..49..137F}. However, we notice that N abundance with twice the solar value matches the predicted N VII line flux better. Earlier, \citet{2017MNRAS.471.2605C} had suggested enhanced N abundance for this source. Hence, we report line ratios for both solar (ratio(1)) and 2$\times$solar abundance (ratio(2)) for Nitrogen in Table \ref{tab:table2}. However, the observed Ne IX line ratio could not be matched even by increasing the Ne abundance. The CLOUDY model also predicts additional, fainter lines beyond those we have detected.

For our model, the dominant heating mechanism is ``H FF", free-free heating. Similarly, total line cooling by O VI,  Ne VII, and Ne IX are the dominant cooling mechanisms \citep{1999ASPC..162.....F}. The thickness of the emitting region is $2\times10^{10}$ cm. However, the model-predicted line intensity for Ne IX is less than the observed value (Fig.~\ref{fig:cloudy_spec}). In the first spectral order, the RGS instruments' resolving power (R) varies from 150 to 800 over a range of 5 to 35 $\AA$ (0.33 to 2.5 keV). To match our observation we create the model spectra for R=150 (Fig.~\ref{fig:cloudy_spec}).

\section{Discussion}\label{disc}  
The spectral characteristics can be used to probe and characterize the circumstellar material and its ionization state in a neutron star LMXB. The O VIII Ly$\alpha$, O VII, N VII, Ne IX (1-2), and Ne IX (1-3) lines are detected in the {\it XMM-Newton}/RGS spectra from EXO 0748--676. The observed emission lines are broadened but do not show any significant Doppler shift.
Thus, it can be inferred that there is no signature of inflow or outflow but rather the material orbits around the central source. 
A discussion on such a plasma, characterized by similar spectral lines, during the previous outburst of this source can be found in \citet{cottam2001high}.

The three O VII lines, i.e., the x+y intercombination line (21.8 {\AA}) and the z forbidden line (22.1 {\AA}), can be used as indicators to identify the source of emission. The intercombination line is detected in all the segments, whereas the forbidden line is absent. This indicates that there is either a high-density photoionized gas or a collisionally ionized gas. 
However, both types of gas can have O VII and O VIII and may show the detected lines. 
Thus, it is hard to disentangle between these two scenarios.

Two different Ne IX emission features are detected during the pre-burst and post-burst segments. They can both be due to different excitation levels of Ne IX. The feature in the pre-burst and total spectra is at 11.65 {\AA} n=1-3, and the feature in the post-burst is at 13.5 {\AA} n=1-2. 
This suggests an impact of the thermonuclear burst on the plasma components, perhaps due to intense radiation.
The Ne IX features are broad and have large FWHM values. \citet{van2009properties} also found a maximum FWHM of 0.35 {\AA}, similar to this work.

The dip and non-dip phase spectra also show a significant impact of dips on the spectral features.  The high-resolution spectra show the difference between these phases. In the dip spectra, the Ne IX (1-2) line is present (unlike the non-dip phase) at 13.53{\AA}. Also, the intensity of the intensity of the O VIII line decreased during the dip phase compared to the non-dip phase (see Table~\ref{tab:line_table}). A decrease in the intensity of the O VIII line can suggest that some of the O VIII emission is absorbed during the intensity dip phases, because during the dips the neutron star is believed to be obscured by a thickened region of the accretion disk \citep{trigo2006spectral}. 

The plasma modeling with (\texttt{CLOUDY}) gives insight into the atmosphere near the central source. The strong emission from both the O VIII Ly $\alpha$ 19{\AA} and O VII  (21.8{\AA}) lines are modeled along with the Ne and N lines. We have also predicted the existence of other fainter lines, not detected in this observation (Table~\ref{tab:table2}). The spectral modeling of EXO 0748--676 using CLOUDY reveals key physical characteristics of the ionized plasma surrounding the source. The model grid best resembling the observed spectrum, places the gas cloud at an inner radius of approximately $3.16\times10^{10}~\mathrm{cm}$ from the neutron star, with an emitting region thickness of 2$\times$10$^{10}$ cm. The electron temperature, calculated to be around 2.13$\times$10$^5$ K, suggests a highly ionized, hot, dense environment. The simultaneous presence of both O VII recombination lines and O VIII emission lines indicates a region with a complex ionization structure, potentially where zones of differing effective ionization overlap due to variations in density or radiation flux. The predicted high value of density can account for the enhanced recombination process and high intensity of the O VII line in all phases. Interestingly, while emission from Ne IX is predicted, the observed Ne IX line intensity could not be reproduced by the model, likely indicating a separate origin for this feature. The Ne IX line’s distinct velocity broadening compared to other lines supports the idea that it may arise from a different spatial region, possibly closer to the neutron star. An enhanced nitrogen abundance, modeled at twice the solar value, produces better matches for the observed N VII line fluxes, suggesting potential nitrogen enrichment in this system. This enrichment could originate from nucleosynthesis processes in the donor star, reflecting a history of CNO cycle processing \citep{jimenez2005identification}.  

In addition to high-resolution spectroscopy, other aspects of this observation are also explored to compare with previous studies. The MOS continuum spectrum is fitted with a power law and a Raymond-Smith (RS) thermal emission model (Appendix~\ref{mos_spec}), similar to \citep{bonnet2001eclipsing}'s EPIC spectral analysis. The RS model is required to fit the sharp rise in the soft energy ($<$2 keV) range. The best-fit parameter values are similar to those from the previous study (Table~\ref{tab:ccd_fit}).
Using the light curves from the soft (0.3--2 keV) and the hard (2--10 keV) bands, the spectral state of EXO 0748--676 is observed (Appendix~\ref{hardness}) to be similar to that reported for the previous outburst \citep{homan2003xmm}.
One type I burst is observed in both X-ray (EPIC) and optical (OM) bands, and the latter lags the former by 3 s (Appendix~\ref{xrayopt}).
The optical burst originates from reprocessed X-rays by the material in the vicinity of the neutron star. 
The lag is a combined result of the light travel time and the radiative reprocessing time. 
Our estimated lag is similar to those reported earlier for the previous outburst of the source
\citep{paul2012simultaneous,hynes2006multiwavelength}. 

Comparing the results from this observation with previous studies on various physical properties of this source reveals striking similarities. The detection of similar elements and atomic transitions in the spectra, the presence of two distinct plasma components, and comparable luminosity and source states indicate that the accretion process remains consistent across outbursts. Observed properties such as the burst lag also align with previous findings, reflecting stability in the burst's temporal properties. However, new results highlight substantial differences too. The impact of bursts on Ne IX emission, as well as changes in line intensities during dip phases, provide new insights into the source's surroundings and the accretion disk's structure.

In summary, while the fundamental characteristics of EXO 0748-676 remain consistent, the new observation provides a better understanding of the system's dynamics and interactions, particularly its response to bursts and intensity dips. 

\section{Conclusion}\label{conc}

This work presents the high-resolution spectral analysis of the neutron star LMXB system EXO 0748--676. We use a single observation from {\it XMM-Newton}, focusing mainly on the RGS data. 
Although EXO 0748--676 has recently gone into an outburst after a long quiescence of 16 years, it shows a striking similarity with the previous outburst. 
Here is a summary of the main findings.
\begin{itemize}
    \item The source is in the same spectral state as it was during the last outburst \citep{homan2003xmm}.
    \item A type I X-ray burst and its optical counterpart have been detected. The latter lags the X-ray burst by 3 s indicating a reprocessing of the burst X-rays.
    \item The total observation is divided into total (burst excluded), pre-burst and post-burst segments. The strongest O VII x,y spectral emission line is detected in all three segments. We have also used a robust Gaussian scanning technique (Appendix~\ref{gauss_scan}) to detect lines of O VIII, Ne IX, and N VII.
    \item The Ne IX line is observed at 11.65{\AA} in the pre-burst and total burstless spectra, and at 13.56{\AA} in the post-burst phase. This suggests a change in the ionization state of the extended plasma region due to the burst.
    \item The Ne IX 13.5{\AA} line is absent in the non-dip phase and the O VIII 19{\AA} line exhibits a decreased intensity in the dip phase.
    \item This work includes the first-ever CLOUDY modeling of EXO 0748–676. The photoionized plasma is found to have a complex ionization structure in which both O VII recombination and O VIII emission lines arise.
    \item Although, the model predicts faint Ne IX 11.65{\AA} emission, the observed Ne IX line's intensity and velocity broadening could not be reproduced by the same plasma. This indicates that the line originates from a separate spatial region closer to the neutron star.
\end{itemize}

This work reports the first detailed results on the spectral lines and other properties of EXO 0748--676 for its second observed outburst after a long quiescence.

Future observations using \textit{XRISM}, \textit{Chandra}, and \textit{XMM-Newton} during the current outburst will be necessary to unveil new features and for a detailed comparison with the previous outburst.

\textit{Acknowledement: }This research is based on observation obtained with XMM-Newton, an ESA science mission with instruments and contributions directly funded by ESA Member States and NASA. This research has also made use of software \citep[HEAsoft]{2014ascl.soft08004N} provided by the High Energy Astrophysics Science Archive Research Center (HEASARC), which is a service of the Astrophysics Science Division at NASA/GSFC. GS acknowledges the WOS-A grant from the Department of Science and Technology (SR/WOS-A/PM-2/2021). The authors also acknowledge the help from \xmm project scientist Dr. Norbert Schartel for his prompt action in obtaining the data through DDT time, and thank an anonymous referee for constructive comments which improved the paper.

\bibliography{references}{}
\bibliographystyle{aasjournal}

\begin{appendices}

\section{Detection of Faint Spectral lines Using Gaussian Scanning}\label{gauss_scan}
\setcounter{table}{0}
\setcounter{figure}{0}
\renewcommand{\thetable}{A\arabic{table}}
\renewcommand{\thefigure}{A\arabic{figure}}
In the RGS spectrum, the O VII line is most prominent whereas other lines are fainter and relatively less significant above the continuum. Previous studies \citep{cottam2001high,van2009properties,psaradaki2018modelling} have shown the detected lines of OVIII, N VII, and Ne IX, X. Hence we try to look for those lines and any other new ones even though this observation has less exposure. 
We confirm the presence of any line (absorption/emission) in the 5 -- 32 \AA~spectrum by fitting a Gaussian with a width increment of 0.05\AA. Then the significance of each line is detected by an F-test by fitting two models (1. just the continuum and 2. continuum with Gaussian line) and using their $\chi^2$ values. Then, the p values from the F-test are used to compute the significance of line detection. Similar to the method of \cite{pinto2016resolved} and \cite{psaradaki2018modelling}, we vary the initial width, and confirm that this variation does not affect the line detections. The significance level for the total burstless spectrum is indicated in Fig.~\ref{gaus_scan}. This process has been carried out for each spectrum and the detection has been reported in Table~\ref{tab:line_table}. It should also be noted that there are other statistically significant emission/absorption features that we couldn't associate with any element's line neither using the CLOUDY model nor using the AtomDB database. A similar situation can be noticed for \citet{psaradaki2018modelling}'s figure 3, especially the absorption lines near O VII and emission lines near Ne IX and N VII. They mentioned those features might be the results of bad pixels.
\begin{figure*}[h]
    \centering
    \includegraphics[width=\textwidth,height=0.6\textwidth]{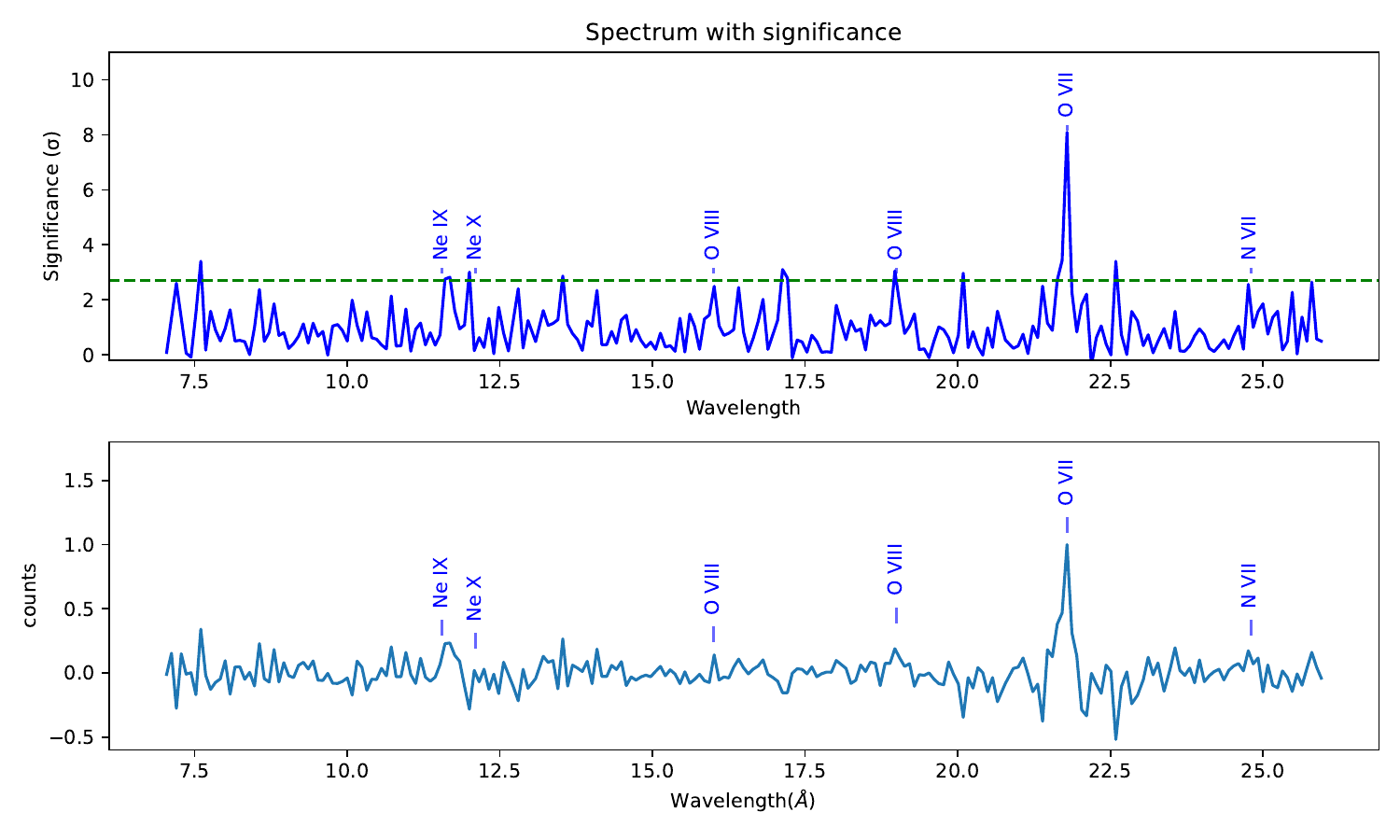}
    \caption{(top) The line significance is shown for the \xmm RGS total spectrum. The horizontal dashed line shows a 2.8 sigma detection level above the continuum to account for the N VII line. (bottom) The \xmm RGS total spectrum with the line IDs for reference (see Appendix~\ref{gauss_scan}).}
    \label{gaus_scan}
    
\end{figure*}

\section{Spectral analysis of MOS CCD spectrum}\label{mos_spec}
\setcounter{table}{0}
\setcounter{figure}{0}
\renewcommand{\thetable}{B\arabic{table}}
\renewcommand{\thefigure}{B\arabic{figure}}

Although the main focus of this work is the high-resolution spectroscopy of EXO 0748--676 using {\it XMM-Newton}/RGS, the combined MOS spectra (1+2 using \texttt{epicspeccombine}) are also used to characterize the continuum. The spectrum is fitted with a simple \texttt{cutoffpl*pcfabs} model in \texttt{XSPEC 12.14}.

\begin{figure*}[h]
    \centering
    \includegraphics[width=\textwidth,height=0.6\textwidth]{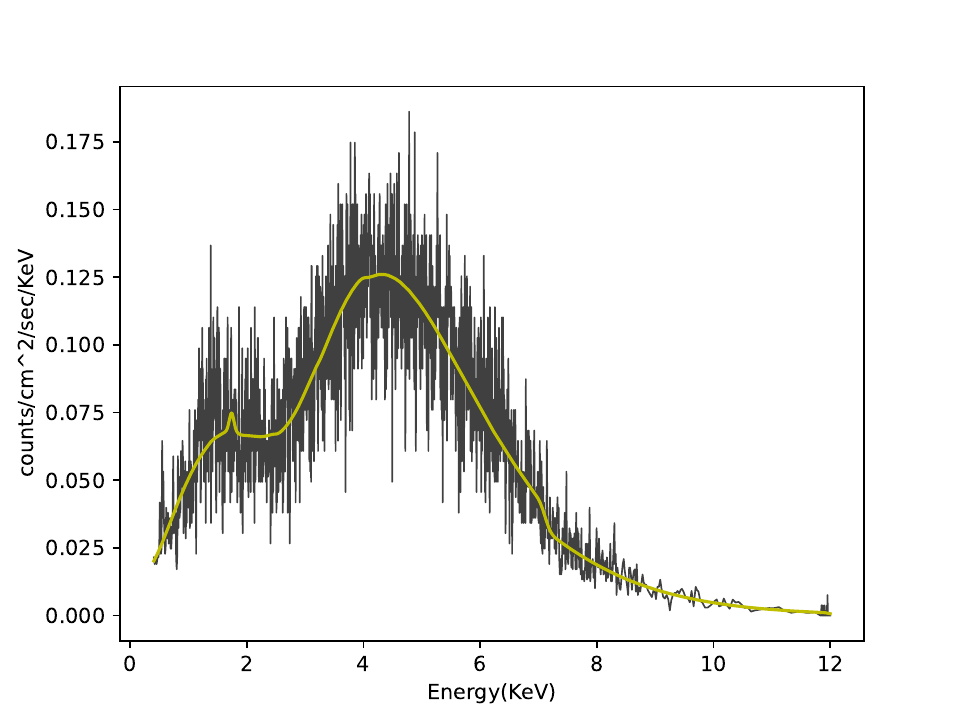}
    \caption{Spectral fit of the total \xmm MOS spectrum of EXO 0748--676 with a \texttt{cutoffpl*pcfabs} model. The spectral data (black) with the fit (yellow) overlayed (see Appendix~\ref{mos_spec}).}
    \label{ccd_spec}
\end{figure*}

The spectrum shows a complex profile in the soft energy range ($<$ 2 keV) that is not fitted with a single power-law model. There is a soft excess evident in the spectrum (Fig. \ref{ccd_spec}), which has also been reported in earlier studies \citep{sidoli2005broad,thomas1997discovery,schulz1999rosat} and the latest telegram on \textit{NuSTAR} observation \citep{buisson2024updated}. Hence, a cutoff power law has been used to fit the continuum with partial absorption. A \texttt{diskbb+powerlaw} model with partial absorption was also used, but it did not result in a good fit ($\chi^2$/dof = 5392.65/1471). The best-fit parameters from the \texttt{cutoffpl*pcfabs} model are given in Table~\ref{tab:ccd_fit}. The parameters were left free for fitting and the resulting fit shows a negative value for the power law index, $\Gamma$ (Table~\ref{tab:ccd_fit}). Although this may not represent the physical process resulting in the continuum shape, it can be used to phenomenologically model the continuum and study the spectral lines using the continuum-normalized RGS spectra. Previous observations during 2000 and 2003 were also fitted with the same model to compare the flux of this source. For both years 2000 and 2003, the observation with the longest duration was reduced and the MOS spectra are fitted with the \texttt{pcfabs*cutoffpl} model to calculate the 0.2-10 KeV flux using \texttt{XSPEC}. The model flux for 2000 and 2003 were respectively 3.37$\times10^{-9}$ ergs/cm$^2$/s and 6.38$\times10^{-9}$ ergs/cm$^2$/s.  

\citet{bonnet2001eclipsing} modeled the EPIC continuum spectra with an additional Raymond-Smith (RS) thermal emission model, along with a power law component, to fit the sharp rise in the soft ($<2$ keV) range. In this work, we also use the same model to fit the MOS spectrum and compare the best-fit parameter values with those of the previous work. 
\citet{bonnet2001eclipsing} created EPIC spectra for flaring, quiescent, and dip phases and jointly fitted using the \texttt{(powerlaw+raymond)*wabs} models. The joint fit values for photon index ($\Gamma$) and plasma temperature (kT) were found to be 1.3 and 0.6 keV (compared to $\approx 1.7$ and $\approx 0.9$~keV, respectively, in this work; see Table~\ref{tab:ccd_fit}), respectively. The better fit obtained by the RS model suggests that the soft X-ray ($<2$ keV) spectrum is a thin hot plasma surrounding the central source. 

We also analyzed the pre- and post-burst MOS spectra to look for potential differences in spectral hardening above 7-8 keV, as this could probe varying excitation levels across these phases. This variation can be linked to the change in excitation levels of the Ne IX ion in pre and post-burst phases. However, our analysis revealed no significant change in the power-law index, indicating that the spectral continuum remains similar before and after the burst. This suggests that any changes in ionization level may not be directly linked to fluctuations in the continuum within the limits of our data. The limited spectral range of XMM-Newton (up to $\sim$10 keV) restricts the degree to which we can probe the spectral changes.

\begin{table*}[tbh]
\caption{Best-fit parameter values using \texttt{cutoffpl*pcfabs} model (top) and \texttt{(powerlaw+raymond)*wabs} model (bottom) for the total \xmm MOS spectrum of EXO 0748--676 (see Appendix~\ref{mos_spec}).}
\centering
    \begin{tabular}{ccc}\label{tab:ccd_fit}
         Model & Parameter & Value  \\ \hline \hline
          & $\Gamma$& -1.81$\pm$0.09 \\
         cutoffpl & HighECut & 1.03 $\pm$ 0.03 keV\\
          & $N_{\rm H}$ & 19.1$\pm$0.9$\times10^{22}$ cm$^{-2}$\\
         pcfabs & coverfrac &0.913$\pm$0.008 \\
          & Flux & (4.0 - 4.15)$\times10^{-9}$ ergs/cm$^2$/s\\
          & ${\chi}^2$/dof & 446.40/1471\\ 
          \hline \hline
          powerlaw & $\Gamma$& 1.68$\pm$0.04\\
          wabs & $N_{\rm H}$ & 12.3$\pm$0.6$\times10^{22}$ cm$^{-2}$\\
          raymond & kT & 0.94$\pm$0.01 keV \\
          raymond & abundance & 1.0 (frozen)\\
          raymond & redshift & 0 (frozen)\\
          & ${\chi}^2$/dof & 2771.78/1471\\  \hline \hline \\
          \end{tabular}
\end{table*}

\section{Evolution of spectral hardness}\label{hardness}
\setcounter{table}{0}
\setcounter{figure}{0}
\renewcommand{\thetable}{C\arabic{table}}
\renewcommand{\thefigure}{C\arabic{figure}}

\begin{figure*}[h]
    \centering
    \includegraphics[width=\textwidth,height=0.6\textwidth]{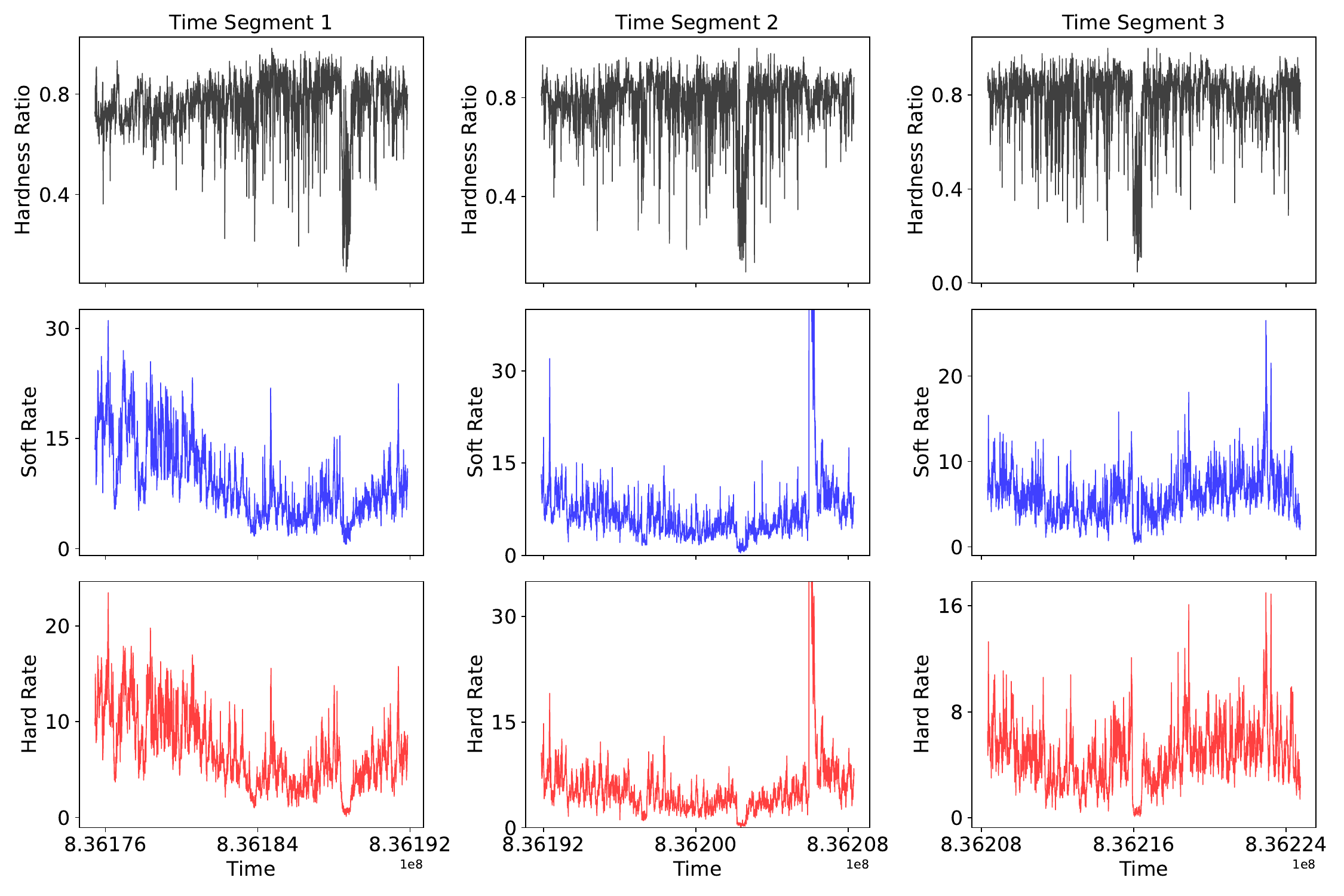}
    \caption{The \xmm PN timing mode light curve of EXO 0748--676 is divided into three equal segments. In each segment, the evolution of the hardness ratio (top row), the soft (0.3--2 keV) light curve (middle row), and the hard (2--10 keV) light curve (bottom row) are shown. 
    The eclipses are more prominent in the hard band but also detected in the soft band (see Appendix~\ref{hardness}).}
    \label{hid}
\end{figure*}

In order to interpret the spectral results and to facilitate a meaningful comparison with past and future observations, it is important to probe the spectral state and spectral evolution of EXO 0748--676 throughout the observation. 
The PN timing mode light curve is used in broad (0.3--10 keV), soft (0.3--2 keV), and hard (2--10 keV) energy bands for this purpose. 
These energy bands are the same as in \citet{homan2003xmm} to compare with the source state during the previous outburst. 
The hardness ratio (HR) is calculated as the ratio of the background-subtracted count rate in 2--10 keV to that in 0.3--10 keV.
A multi-panel diagram (Fig.~\ref{hid}) shows the time evolution of the HR, the 0.3--2 keV count rate, and the 2--10 keV count rate.
 
The source remains in the same state for the entire duration of the observation (mean HR $\approx 0.77$). The spectral state is softer during the eclipse compared to the average. Although hardness ratio values are not explicitly reported, Fig. 1 of \citet{homan2003xmm} shows that the characteristics of the source reported here are similar to those observed earlier. Hence, it can be seen that the source has returned to its earlier spectral state after about 21 years.

\section{CLOUDY Model: Figures and Table}\label{cloudyfigtab}
\setcounter{table}{0}
\setcounter{figure}{0}
\renewcommand{\thetable}{D\arabic{table}}
\renewcommand{\thefigure}{D\arabic{figure}}
{Here, the parameters of the CLOUDY model are shown in Table~\ref{tab:table1}. The observed and predicted line ratios are compared in Table.~\ref{tab:table2} and Fig.~\ref{fig:cloudy_spec}.} 
\begin{table*}[t]
	\centering
	\caption{Predicted model parameters using CLOUDY.}
	\label{tab:table1}
	\begin{tabular}{lr} 
		\hline
		Physical parameters & Predicted values\\
		\hline
		Power law: $\alpha$, luminosity (erg s$^{-1}$) &  -1.44, 10$^{35}$\\
                T$_{BB}$ (K) , luminosity (erg s$^{-1}$) & 10$^{7}$, $3.16 \times 10^{36}$ \\
		Density $n_{\rm H}(r_{0})$(cm$^{-3}$) & $3.98 \times 10^{13}$\\
		Inner radius r$_{0}$ (cm) & $3.16 \times 10^{10}$ \\
        Thickness of the gas (cm) & $2.04 \times 10^{10}$\\
        Covering factor & 0.3\\
		\hline				
	\end{tabular}
\end{table*}

\begin{table*}[t]
	\centering
	\caption{Comparison of observed (total spectrum) and predicted line ratios with respect to the OVIII (18.9689 $\AA$) line.}
	\label{tab:table2}
	\begin{tabular}{lllllll}
\hline
	Observed & $\lambda$ & Observed & Predicted & $\lambda$(rest frame) & Predicted & Predicted\\          
      lines\footnote{These predicted lines, while fainter than the O lines are seen to be stronger than the Ne IX 11.55 line, although not detected in our current spectra, they might get detected in a future observation.} & ({\AA}) & ratio\footnote{The ratio is calculated with respect to the intensity of the O VIII line.} & lines & ({\AA}) & ratio (1)\footnote{These ratios are predicted by the model assuming solar abundance for nitrogen.} & ratio (2)\footnote{These ratios are predicted by the model assuming twice the solar abundance for nitrogen.}\\ 
		\hline
		Ne IX & 11.65$\pm$0.03 & 0.503 & Ne IX & 11.5466 & 0.024 & 0.028\\
		O VIII & 18.99$\pm$0.03 & 1.000 & O VIII & 18.9689 & 1.000 & 1.000 \\
		O VII & 21.79$\pm$0.02 & 2.149 & O VII(Blnd) & 21.8070 &2.013 & 2.311 \\
		N VII & 24.81$\pm$0.04& 0.248 & N VII & 24.7810 & 0.138 & 0.249\\
         &   &  & Mg XI(Blnd) & 9.23121 & 0.078 & 0.088\\
         &   &  & Ne IX(Blnd) & 13.5529& 0.526 & 0.611\\
         &  &  & Ne X & 12.1339& 0.032 & 0.035\\
         &  &  & O VIII & 16.0059 & 0.011 & 0.012\\
         & & & O VII & 17.7680 &0.020 & 0.020\\
         & & & O VII & 18.6270 &0.033 & 0.036\\
         &  &  & C VI & 28.4656 & 0.094 & 0.109\\
         &  &  & N VI & 28.7870 & 0.193 & 0.420\\
         & & & N VI(Blnd) & 29.084 &0.601 & 1.501\\
		\hline
	\end{tabular}
\end{table*}

\begin{figure*}
\includegraphics[height=0.5\textwidth, width=0.5\textwidth]{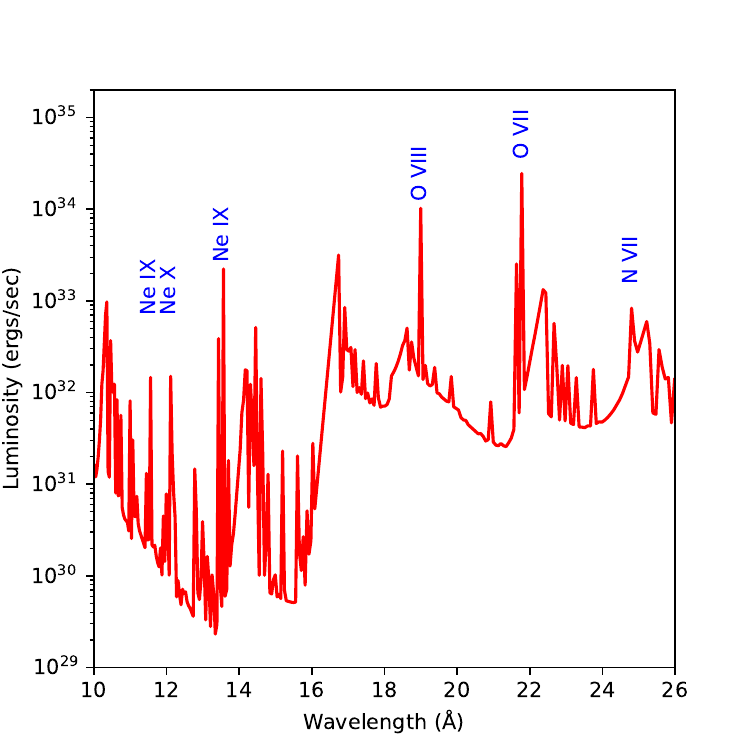}
\includegraphics[height=0.5\textwidth, width=0.55\textwidth]{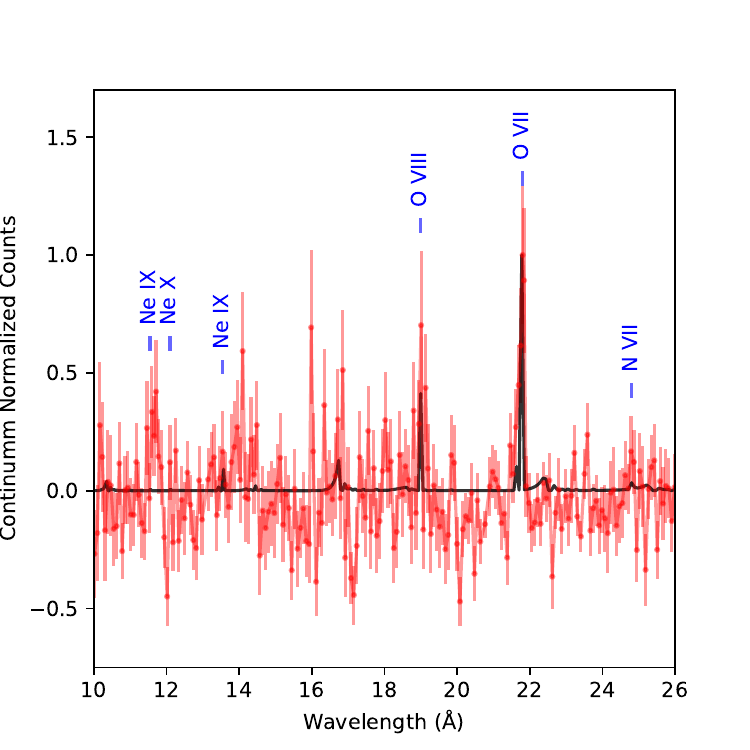}
\caption{(left) CLOUDY model spectrum showing the expected lines. (right) Model spectrum overlayed on the observed spectrum, showing the detected lines.}
\label{fig:cloudy_spec}
\end{figure*}

\section{X-ray and Optical bursts}\label{xrayopt}
\setcounter{table}{0}
\setcounter{figure}{0}
\renewcommand{\thetable}{E\arabic{table}}
\renewcommand{\thefigure}{E\arabic{figure}}
 
At the time of the observed thermonuclear X-ray burst from EXO 0748--676, we also detect an optical burst from the {\it XMM-Newton} Optical Monitor (OM) data.
We use the {\it XMM-Newton} EPIC/MOS light curve and the OM fast mode light curve of EXO 0748--676, both binned at 0.5 s (the timing resolution of OM fast mode), to compare the bursts in X-ray and optical bands (170--650 nm). The fast mode data were acquired for an approximately 22$\times$23 pixel square window at a high timing resolution (0.5 seconds).
\begin{figure*}[h]
    \centering   \includegraphics[width=0.55\textwidth,height=0.5\textwidth]{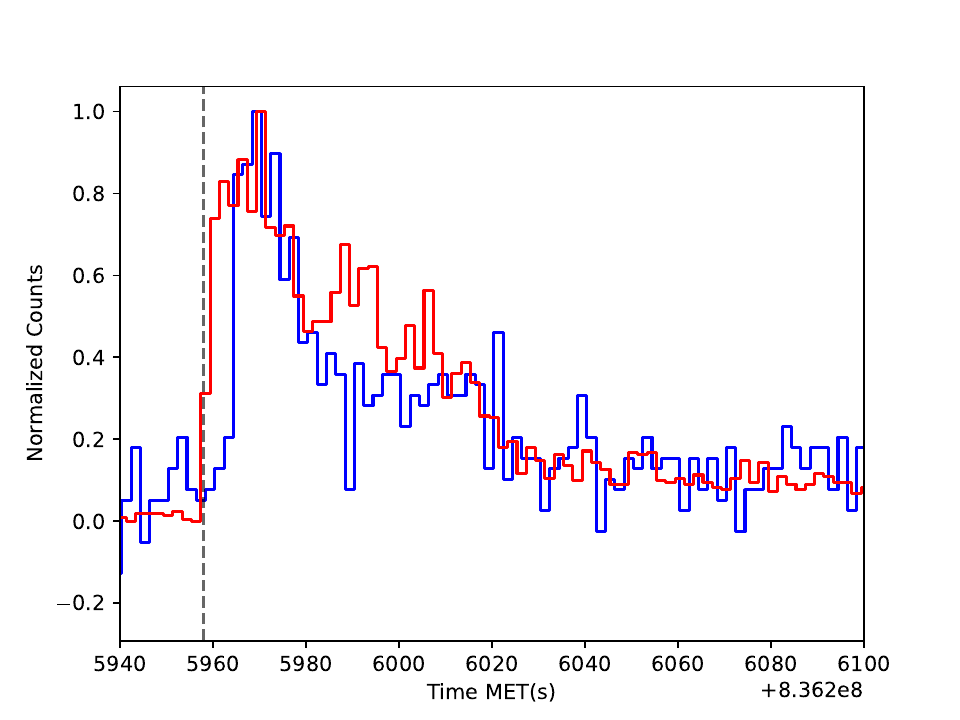}
    \includegraphics[width=0.35\textwidth,height=0.45\textwidth]{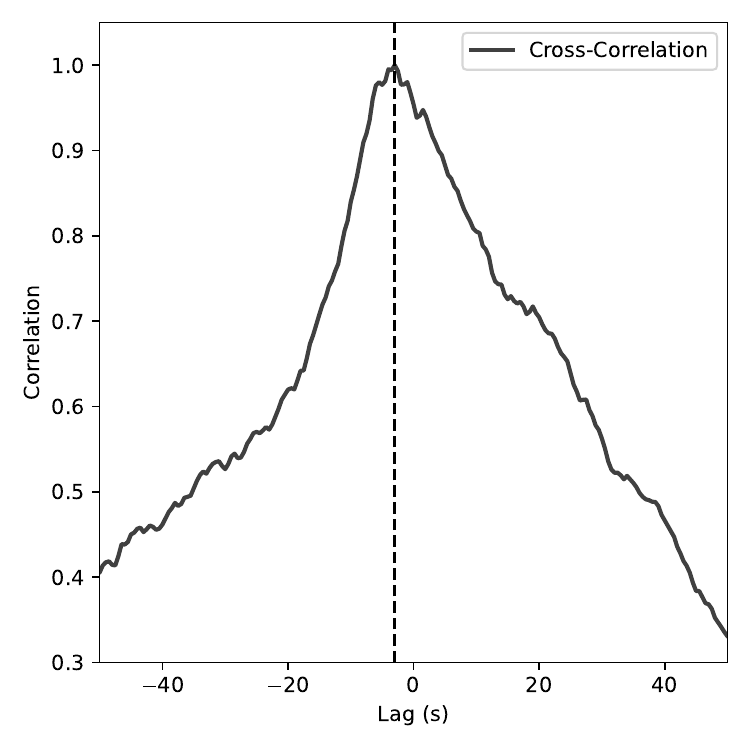}
    \caption{Left: \xmm  OM fast mode light curve (blue) and MOS light curve (red) of EXO 0748--676. 
    Both are binned at 2 s for visualization, and the vertical dashed line is placed at 836205958 s MET to mark the rise time of the X-ray burst. 
    Right: Cross-correlation plot between the X-ray and optical light curves. The vertical dashed line marks a 3-second lag of the optical burst with respect to the X-ray burst
    (see Appendix~\ref{xrayopt}).}
    \label{om_pn}
\end{figure*}
The MOS and OM barycenter-corrected light curves are binned at 2 s for visual inspection (Fig.~\ref{om_pn}) and 0.5 s for quantitative determination of the lag between them. 

The X-ray and optical light curves are cross-correlated using a method similar to that described in \citet{hynes2006multiwavelength}, and it is estimated that the optical bursts lag the X-ray burst by 3 s.
This value is similar to a previously found average lag of 3.25 s \citep{paul2012simultaneous}.

A significant number of optical photons detected from LMXBs is due to the reprocessing of X-rays from the circumstellar material.
Thus, a thermonuclear burst is expected to manifest in the UV/optical light curves as well \citep{paul2012simultaneous, hynes2006multiwavelength}.
An optical burst can occur due to the reprocessing of X-rays by material near the neutron star within a few light seconds (light travel time), and this information can be used to probe the accretion components, size, and geometry around the compact object. 
This is because the observed delay in the optical burst after an X-ray burst should depend on the position and nature of the reprocessing region and the geometry of the binary system. 
An optical burst could also originate from the X-ray reprocessing from the surface of the companion star \citep{hynes2006multiwavelength}. 
However, \citet{paul2012simultaneous} found no evidence of orbital modulation of these features using 63 simultaneous X-ray and optical bursts from EXO 0748--676. 
The delay in X-ray and optical bursts can be due to a combination of light travel time and radiative reprocessing time.

\end{appendices}
\end{document}